\documentstyle[aasms4,epsf]{article}

\begin{document}
\def\rpcomm#1{{\bf COMMENT by RP:  #1} \message{#1}}
\def\uptilde{\mathaccent"164}
\def\etal{{\it et al.~}}
\def\ls{\vskip 12.045pt}   
\def\ni{\noindent}        
\def\kms{km\thinspace s$^{-1}$ }     
\def\amm{\AA\thinspace mm$^{-1}$ }     
\def\deg{\ifmmode^\circ _\cdot\else$^\circ _ \cdot$\fi }    
\def\degg{\ifmmode^\circ \else$^\circ $\fi }
\def\solar{\ifmmode_{\mathord\odot}\else$_{\mathord\odot}$\fi} 
\def\arcs{\ifmmode {'' }\else $'' $\fi}     
\def\arcm{\ifmmode {' }\else $' $\fi}     
\def\buildrel#1\over#2{\mathrel{\mathop{\alphall#2}\limits^{#1}}}
\def\mper{\ifmmode \buildrel m\over . \else $\buildrel m\over .$\fi}
\def\hper{\ifmmode \rlap.^{h}\else $\rlap{.}^h$\fi}
\def\sper{\ifmmode \rlap.^{s}\else $\rlap{.}^s$\fi}
\def\arcsper{\ifmmode \rlap.{' }\else $\rlap{.}' $\fi}
\def\arcmper{\ifmmode \rlap.{'' }\else $\rlap{.}'' $\fi}
\def\gapprox{$_ >\atop{^\sim}$}     
\def\ltapprox{$_ <\atop{^\sim}$}     
\def\tworule{\noalign{\medskip\hrule\smallskip\hrule\medskip}}
\def\onerule{\noalign{\medskip\hrule\medskip}}
\def\et{{\it et~al.~}}
\newcommand{\lta}{{\small\raisebox{-0.6ex}{$\,\stackrel
{\raisebox{-.2ex}{$\textstyle <$}}{\sim}\,$}}}
\newcommand{\gta}{{\small\raisebox{-0.6ex}{$\,\stackrel
{\raisebox{-.2ex}{$\textstyle >$}}{\sim}\,$}}}
\newcommand{\apb}{$(A+B)/2$ }
\newcommand{\amb}{$(A-B)/2$ }

\def\apj{ApJ}  
\def\apjs{Ap.~J.~Suppl. }  
\def\apjl{Ap.~J.~ } 
\def\pasp{{Pub.~A.S.P.} }      
\def\mn{MNRAS}      
\def\aa{Astr.~Ap. }     
\def\aasup{AAS }     
\def\baas{Bull.~A.A.S. }  

\lefthead{Guti\'errez et al.}
\righthead{The Tenerife experiments}

\title{THE TENERIFE COSMIC MICROWAVE BACKGROUND MAPS: OBSERVATIONS AND FIRST ANALYSIS}

\author{C. M. Guti\'errez\altaffilmark{1}, R. Rebolo\altaffilmark{1,2}, R. A. Watson\altaffilmark{1,3}, \\R. D. Davies\altaffilmark{3},
A. W. Jones\altaffilmark{4}, and A. N. Lasenby\altaffilmark{4}}

\altaffiltext{1}{Instituto de Astrof\'\i sica de Canarias, E~38200 La Laguna, Tenerife, Canary Islands, Spain}

\altaffiltext{2}{Consejo Superior de Investigaciones cient\'\i ficas}

\altaffiltext{3}{Mullard Radio Astronomy Observatory, Cavendish Laboratory,
	Madingley Road, Cambridge CB3 OHE, UK}

\altaffiltext{4}{University of Manchester, Nuffield Radio Astronomy
	Laboratories, Jodrell Bank, Macclesfield, Cheshire SK11 9DL, UK}

\begin{abstract}

The results of the Tenerife Cosmic Microwave Background (CMB)
experiments are presented. These observations cover 5000 and 6500
square degrees on the sky at 10 and 15 GHz respectively centred around
Dec.$\sim +35\degg$. The experiments are sensitive to multipoles
$l=10-30$ which corresponds to the Sachs-Wolfe plateau of the CMB power
spectra. The sensitivity of the results are $\sim 31$ and $\sim 12$
$\mu$K at 10 and 15 GHz respectively in a beam-size region ($5\degg $
FWHM). The data at 15 GHz show clear detection of structure at high
Galactic latitude; the results at 10 GHz are compatible with these, but
at lower significance. A likelihood analysis of the 10 and 15 GHz data
at high Galactic latitude, assuming a flat CMB band power spectra gives
a signal $\Delta T_\ell=30^{+10}_{-8}$ $\mu$K (68 \% C.L.). Including
the possible contaminating effect due to the diffuse Galactic
component, the CMB signal is $\Delta T_\ell=30^{+15}_{-11}$ $\mu$K.
These values are highly stable against the Galactic cut chosen.
Assuming a Harrison-Zeldovich spectrum for the primordial fluctuations,
the above values imply an expected quadrupole
$Q_{RMS-PS}=20^{+10}_{-7}$ $\mu$K which confirms previous results from
these experiments, and which are compatible with the COBE DMR data in
the case of the standard inflationary Cold Dark Matter models.

\keywords{cosmic microwave background-cosmology: observations}
\end{abstract}

\newpage

\section{INTRODUCTION}

In the last five years, experiments on angular scales from several
degrees to a few minutes have started to delineate the CMB power
spectra (see Smoot (1997) for a recent review), offering a unique view
in astronomy on the conditions prevailing in the early Universe and the
opportunity to determine the most important cosmological parameters.
The current observations put constraints on the level of normalization
of the Sachs-Wolfe effect, and appear to establish the existence of the
first acoustic peak (Hancock \et 1998). However, the small amplitude of
the CMB anisotropy ($\Delta T/T\sim 10^{-5}$) and the necessity to
measure it within a precision of $\Delta T/T\sim 10^{-6}$ to determine
the cosmological parameters with uncertainties $\simeq 10 \%$, is in
the limit of the sensitivity of current technology and a challenge for
the next generation of CMB experiments.  Here, we present the results
of the Tenerife experiments at 10 and 15 GHz in a sky region covering
$\sim 5000$ and $\sim 6500$ square degrees respectively of which we
have selected $\sim 2000$ square degrees lying at high Galactic
latitude. The multifrequency nature of the experiment allows a
separation of possible contributions of the diffuse Galactic emission
and other foregrounds from the CMB fluctuations. The observations
presented here are sensitive to scales $\sim 2\degg-6\degg$ (the
Sachs-Wolfe regime of the CMB power spectra). This angular range can be
used to test the inflationary models, to measure directly the level of
normalization, the spectral index of the primordial fluctuations and
the possible existence of a foreground of gravitational waves. The
paper is organized as follows:  Section~2 presents a brief summary of
the observations and data processing, Section~3 presents the maps of
the sky at both frequencies.  The contribution of foregrounds is
considered in Section~4. Section~5 analyses statistically the
properties of the signals detected, while Section~6 summarises the
results and brings together the conclusions reached.

\section{OBSERVATIONS}

The Tenerife CMB experiments are a collaboration between the University
of Manchester, the Instituto de Astrof\'\i sica de Canarias (IAC) and the Mullard Radio Astronomy Observatory (MRAO)  which started in
the year 1984 with the objective to detect and map the CMB structure
generated by the Sachs-Wolfe effect. The observations are conducted by
three drift scanning radiometers at 10, 15 and 33 GHz installed at The
Teide Observatory in Tenerife (Spain). The instruments have two
independent receiver channels and operate using a double-difference
technique, with a beam response of the form $-$0.5, $+1$, $-$0.5 (the
positive in the meridian and the other two displaced 8\deg 1 in right
ascension) and a primary beam of full width half maximum (FWHM) $\sim
5\degg$.  The instruments are sensitive to a range in multipoles
($\ell\sim 10-30$) which corresponds roughly to angular scales on the
sky of $\theta \sim 2\degg-6\degg$. A detailed description of the
instruments and observing strategy can be found, for example, in Davies
\et (1996).

This paper presents the data taken by the experiments at 10 and 15 GHz
until the end of 1997. The instruments observe strips of the sky
separated by 2\deg 5, this separation (half of the FWHM) fully samples
the region between adjacent declinations allowing the reconstruction of
a map of the sky fluctuations at each frequency. The strips analysed
here range from Decs.=32.5\degg~to 42.5\degg~at 10 GHz, and from
Decs.=+27.5\degg~to +45\degg~at 15 GHz. We have not included data from
the 33 GHz radiometer as the sky coverage is still sparse as compared
to the other two experiments.
Figure~1 presents the sky
region observed in a Mollweide projection; the ring represents full
strips in RA with the inner and outer edges corresponding to
Dec.=+27.5\degg~and +45\degg~respectively (the 15~GHz data).  The
analysis presented here is limited to the section of our data at high
Galactic latitude ($|b|\gta 40$\degg). The pre-processing (not
presented in this paper) includes calibration, discarding data closer
than 50\degg~and 30\degg~to the Sun and Moon respectively  and the
removal of long term (on timescales of several hours) atmospheric
baselines simultaneously with the Maximum Entropy analysis used to
reconstruct the intrinsic sky signal (see Jones \et 1998). The
surviving data in each position of the sky are stacked together in
order to reduce the level of the noise. These stacked second-difference
data at high Galactic latitude at each declination and frequency are
shown in Figure~2. Table~1 presents a summary of the sensitivity
achieved in a beam size (5\degg) and the average number of individual
measurements per pixel contributing in each of these declination
strips. At 10 GHz the coverage is nearly uniform in terms of noise and
number of  observations; however at 15 GHz the coverage is less uniform
with the poor declination being Dec.=+27.5\degg~which is twice as noisy
as the rest of the scans. From the table, it is possible to compute
the mean sensitivity on a single $1\deg$ measurement, obtaining $\sim
480$ and 225 $\mu$K at 10 and 15 GHz respectively. This difference in
sensitivity is intrinsic to the instruments, being mainly a result of
the difference in the bandwidth in frequency of the detectors ($\sim
$0.5 and $\sim$ 1.5 GHz for the receptors at 10 and 15 GHz
respectively).

\begin{table}[thp]
\caption{Sensitivity of the data at 10 and 15 GHz}
\vspace{0.4cm}
\begin{center}
\begin{tabular}{ccccc}
\multicolumn{1}{c}{Dec. ($^{\circ}$)} &\multicolumn{2}{c}{10 GHz} & \multicolumn{2}{c}{15 GHz}\\
\hline
  & Num & $\sigma$ ($\mu$K) & Num. & $\sigma$ ($\mu$K) \\
\hline
27.5 &    &    &  48 & 35 \\
30.0 &    &    & 151 & 18 \\
32.5 & 76 & 58 & 120 & 21 \\
35.0 & 74 & 56 & 160 & 18 \\
37.5 & 81 & 51 &  96 & 21 \\
40.0 & 95 & 50 & 130 & 22 \\
42.5 & 66 & 54 &  84 & 23 \\
45.0 &    &    &  83 & 23 \\
\end{tabular}
\end{center}
\end{table}

\section{MAPS AT 10 AND 15 GHZ}

The data presented in Figure 2 represent the intrinsic signal on the
sky convolved with the triple beam response of the experiment and with
the addition of baselines and noise. To obtain the intrinsic signal on
the sky some method of deconvolution is needed; several techniques have
been applied to astronomical data: CLEAN, the Wiener filter (Bunn,
Hoffman and Silk, 1996) and the Maximum Entropy Method (MEM) (White \&
Bunn (1995), Jones \et 1998). The MEM method has been applied to the
Tenerife CMB data 10 GHz data with the old instrumental configuration
(8\degg~FWHM). The method allows for the reconstruction of positive and
negative features by considering the intrinsic CMB structure as the
difference of two positive additive distributions as explained in
Maisinger, Hobson, \& Lasenby (1997). The details of the method and its
implementation for the analysis of the Tenerife data can be found in
Jones \et (1998). The only differences with that work are the beam
size of the instruments (5\degg~FWHM) instead of 8\degg~FWHM in the old
configuration, and that now the contribution of the point sources is
subtracted in each of the observations before the MEM reconstruction.
The estimation of this contribution is explained in Section~4.

We applied the MEM deconvolution to the 10 and 15 GHz data
independently; an extension of the technique with a joint analysis of
data taken at several frequencies is now in progress and will be
presented in a future paper. The sampling of the observations (1\degg~in
RA and 2.5\degg~in Dec.) is enough to fully sample the sky. The main
feature on the observed sky region is the emission from the Galactic
plane with amplitudes as large as 60 mK for the major crossing at 10
GHz in the strip at Dec.=+40\degg~(Guti\'errez \et 1995). These
amplitudes are several order of magnitudes larger than the expected
amplitudes due to the contribution of discrete radio sources, the
diffuse emission at high Galactic latitudes, and the CMB fluctuations.
As was pointed out in the paper by Jones \et (1998), this large
dynamical range prevents the simultaneous reconstruction of both the
dominant features due to the Galactic plane crossing and the features lying
at high Galactic latitudes.  The reconstruction presented here is
limited to the section of our observations at high Galactic latitudes
$|b|\gta 40$\degg.

Figure 3 shows the maps of the sky reconstruction at 10 and 15 GHz
after subtraction of the discrete point-source contribution (see next
section). The errorbars were calculated as a variance over 300
Monte-Carlo reconstructions of the same sky with different noise
realisations giving values of $46$ $\mu$K and $35$ $\mu$K per
1\degg~square pixel at 10 and 15~GHz respectively, or $\sim 8$ and
$\sim 6$ $\mu$K respectively in a beam-size region ($5\degg\times
5\degg$) (compared with $31$ $\mu$K and $12$ $\mu$K in the raw data).

The two MEM processed maps can be compared to each other to discern the
origin of the fluctuations present. Features present in both the 10~GHz
and 15~GHz reconstructed maps are likely to be CMB in origin whereas
features that appear with a greater amplitude in the 10~GHz
reconstruction are likely to be Galactic in origin. It is important to
note that the greyscales on the two maps are not the same so that a CMB
feature will appear less bright (by a factor of 2) in the 10~GHz map
than in the 15~GHz map.

The two most obvious CMB features that appear in both maps are at Dec.
$40\degg$, RA $180\degg$ and Dec. $35\degg$, RA $220\degg$. Both appear
with an amplitude of about 200 $\mu$K in both maps and are therefore
very likely to be CMB in origin. These are in agreement with the
results of the comparison between the COBE DMR and the Tenerife data
(Lineweaver et al. (1995), Guti\'errez et al. 1997). Possible Galactic
sources (corresponding to either synchrotron or free-free emission)
seen in the 10~GHz map and at about half the amplitude in the 15~GHz
map are at Dec. $40\degg$, RA $210\degg$ and Dec. $37.5\degg$, RA
$240\degg$ (which was also seen at 10~GHz in the 8 degree switched beam
reported by Jones {\em et al.}, (1998)). There are numerous other
features that are unaccounted for which appear in the two maps. Some
may be due to noise or residual point sources but others may be
Galactic or CMB in origin that the present data has not been able to
constrain very accurately. For example, the feature which just appears
at the bottom of the 10~GHz map (Dec. $32.5\degg$, RA $180\degg$) and
appears in the 15~GHz map as an extended source at slightly lower
declination and RA may be Galactic in origin. This source has been
detected in the Jodrell Bank  5~GHz survey (Jones {\em et al.}
(1999)$b$) with an amplitude implying a Galactic spectral index
(between 2 and 3) between 5~GHz and 15~GHz.A more comprehensive joint
analysis of the two data sets allowing a seperation of the CMB and
Galactic components will be presented in a later paper.

\section{FOREGROUNDS}

\subsection{Point sources}

The contribution of the discrete point sources  has been estimated
from the measurements by the Michigan monitoring programme (Aller \&
Aller 1997); these have been complemented by the K\"{u}hr \et (1981)
and Green Bank (Condon, Broderick \& Seielstad 1989) catalogues of discrete radio sources. Our estimation for
this foreground takes into account the time variability of each source
and represents an improvement over previous estimations using just the
K\"{u}hr \et and Green Bank catalogues. Given the beam size of the
instruments (5\degg~FWHM) the contribution of any point like source  is
highly diluted in our data, for instance a source with a flux of 1 Jy contributes 35 and 12
$\mu$K at 10 and 15 GHz respectively. In the section of our data at high Galactic latitudes the main contributors are the sources 3C345
(RA=$16^h41^m18^s$, Dec.=$39\degg54^{\prime}11^{\prime \prime}$), 4C39.25
(RA=$9^h23^m56^s$ Dec.=$39\degg 15^{\prime}23^{\prime \prime}$) and
3C286 (RA=$13^h28^m50^s$, Dec.=$30\degg45^{\prime}59^{\prime \prime}$).
Of the three mentioned sources only 3C286 does not show significant
time variability. The other two sources change over the observing
period by factors of $2-3$, which makes it essential to have
time-dependent data for them. In general, sources with smaller
fluxes also show similar relative time variability, but their effect on
our measurements is not significant.

We have taken the measurements by Aller \& Aller at 4.8, 8.0 and 14.5
GHz; which include all the sources in our observing region with
amplitudes $\gta 1$ Jy. We fitted a polynomial to the flux variability
at each frequency and checked the consistency of this fit with the
predictions (Guti\'errez \et 1999). For each month we computed the
corresponding flux at our observing frequencies which was then
converted into antenna temperature and convolved with the beam of our
experiments to reproduce the contribution of such sources in our data.
The {\em rms} of our data due to the contribution of these discrete
radio sources in the region RA=130\degg$-$260\degg~lies between $\sim
40$ and $\sim 120$ $\mu$K, and between $\sim 15$ and $\sim 40$ $\mu$K
at 10 and 15 GHz respectively in a bin of 1\degg~in RA; the larger rms
occurs in the scan at Dec.=+40\degg~and is due mainly, as has been
explained above, to the contribution of the sources 4C39.25 and 3C345.
The contribution due to a foreground of unresolved radio-sources is
expected to have an rms $\lta 30$ and $\lta $15 $\mu$K at 10 and 15 GHz
respectively (Franceschini \et 1989).

Figure~4 shows the comparison between the predicted contribution of the
main radio sources (3C 345, 3C39.25 and 3C 286) and our measurements.
At 15 GHz the peak contribution of the sources is in the range $70-120$
$\mu$K, while at 10 GHz the range is $300-400$ $\mu$K. The agreement between the
prediction of the contribution of these radio sources and our
measurements represents an important check on the consistency of our
data.

\subsection{Diffuse Galactic component}

The most direct approach to analyse the possible diffuse Galactic
component is by using low frequency surveys, like the ones at 408 MHz
(Haslam \et 1982) and 1420 MHz (Reich \& Reich 1988). However, there are
several papers showing the unreliability of these two surveys  when used to predict this contribution for
CMB experiments working at centimeter wavelengths (for example, Davies,
Watson \& Guti\'errez  1996). Additional problems for these predictions
come from the fact that neither the frequency spectral index  or the spatial power
spectra of the diffuse Galactic emission is well known. Existing
evidence shows a steepening of this frequency spectral index with frequency. For
instance Bersanelli \et (1996) found a value $2.9\pm 0.3$ between 1.42
and 5 GHz at angular scales $\sim 2\deg$.  At larger angular scales
Platania \et (1997) obtained a spectral index for the synchrotron emission near the Galactic plane of
2.81$\pm 0.16$ in the range 1.0-7.5 GHz, with evidence of steepening at
frequencies above 7.5 GHz.

The limitations mentioned above on the low frequency surveys have
motivated the development of alternative methods to separate this
foreground from the CMB structure. One of these will be presented in a
forthcoming paper (Jones \et (1999)$a$ in preparation) in which the
Tenerife data at 10, 15 and 33 GHz, and the COBE DMR data (Bennett \et
1996) are incorporated in an extension of the MEM method reconstructing
the most probable CMB and Galactic maps. Other analyses which
cross-correlate sky surveys at centimeter and millimeter wavelengths
with the data presented here, are also in progress. In this work we
have made a joint likelihood analysis of the data at 10 and 15 GHz.
This method  provides by itself  a statistical way to separate the CMB
and the Galactic component; this is explained in next section.

\section{STATISTICAL ANALYSIS}

\subsection{Likelihood analysis}

In standard inflationary models the CMB fluctuations are a realization
of a random gaussian field, and the probability distribution is fully
determined  by the angular power spectrum or alternatively by the
Legendre transformation, the two-point correlation function. This can
be determined using a likelihood analysis.  Methods based on the
likelihood function have been extensively used in the analysis of CMB
data (see for example Hancock \et (1997) for a general discussion, and
Guti\'errez \et (1995) for the application of the method to the Tenerife
data).  The method computes by a Bayesian inference the statistical
probability distribution of the sky signal, considering the correlation
present between each pair of data points and taking into account the
observing strategy, experimental configuration, properties of the
noise, etc.  Experimental data are the result of convolution of the sky
signal with the beam response of the instruments, with the addition of
some noise. This response can be characterized by the experimental
window function $W_\ell$ which defines the filtering produced by the
experiment at each angular mode of the power spectrum. In the case of
the Tenerife experiments the window function is sensitive to a range in
mutipoles $\ell\sim 10-30$ with a maximum response to $\ell\sim 20$. In
addition to this filtering, the observed correlation function contains
the contribution from the noise.  The likelihood function then follows a
multinormal distribution with the covariance matrix composed of two
terms, one due to the signal and one due to the noise: ${\bf C=S+N}$.
In our case the data consist of a set of second differences in
temperature at 10 and 15 GHz, along with their errors, binned at
1$^{\circ}$ in RA. For the matrix of the noise, the only non-zero terms
are those on the diagonal, as the instrumental noise is uncorrelated in
scales larger than the sampling in RA (Guti\'errez \et 1997). The
method assumes a given model for the sky signal with a free parameter
(the level of normalization) which will be constrained according to its
compatibility with the data.

We have applied the likelihood analysis to the data at 10 GHz and 15
GHz after subtraction of the discrete point-source contribution. We
assumed a model for the sky signal in which the contribution by each
multipole to the total power spectrum is the same for all multipoles at
which the experiment is sensitive; these kind of models are called flat
band power spectrum. We restricted the analysis to sections of data in the observing region  RA=120\degg-270\degg~which is away from the Galactic plane. Table~2 summarizes the results. The quantity
quoted is $\Delta T_\ell=\sqrt{\ell(\ell+1)C_\ell/2\pi}$. The table
also shows the value of the likelihood peak relative to the value in
the absence of signal, which shows the high significance of the
detected signal. We see that including the section
RA=230\degg$-$250\degg~the signal detected is slightly larger. For the
estimation of the CMB signal we are more confident on the results in the
range RA=$161\degg-230\degg$ which excludes possible residuals due to
the contribution of the variable sources 3C345 and 4C39.25.

We have performed the same analysis for the 10 GHz data. In the
analysis of each scan separately there is no clear evidence of
detection in the RA range $120\degg-260\degg$ except for the scan at
Dec.=+32.5\degg~which has a detection of signal at the two-sigma level.
The simultaneous analysis of the 5 scans gives the result shown in
Table~2. The detection here is compatible with the results at 15 GHz,
but  less significant as a consequence of the higher noise and less sky
coverage at each frequency as compared with the data at 15 GHz.

\begin{table}[thp]
\caption{Likelihood results at 10 and 15 GHz}
\vspace{0.4cm}
\begin{center}
\begin{tabular}{lcccc}
\hline
Data &  Dec. (\degg) & RA (\degg) & $\Delta T_\ell$ ($\mu$K)& $L_{max}$\\
\hline
15 GHz    & 27.5-45   & 141-210 & $35^{+8}_{-7}$ & $3.2\times 10^{10}$ \\
15 GHz    & 27.5-45   & 141-230 & $35^{+8}_{-6}$ & $1.3\times 10^{15}$ \\
15 GHz    & 27.5-45   & 141-250 & $39^{+7}_{-7}$ & $2.6\times 10^{29}$ \\
15 GHz    & 27.5-45   & 161-230 & $30^{+9}_{-7}$ & $1.7\times 10^{8}$  \\
15 GHz    & 27.5-45   & 161-250 & $37^{+8}_{-6}$ & $2.1\times 10^{22}$ \\
15 GHz    & 27.5-45   & 181-250 & $32^{+8}_{-6}$ & $1.6\times 10^{19}$ \\
\hline
\hline
10 GHz    & 32.5-42.5 & 141-210 &  $\le 47$ & 1.6  \\
10 GHz    & 32.5-42.5 & 141-230 &   $\le 35$ & 1.1 \\
10 GHz    & 32.5-42.5 & 141-250 &  $29^{+10}_{-10}$ & 57.0 \\
10 GHz    & 32.5-42.5 & 161-230 &  $\le 30$ & 1.0 \\
10 GHz    & 32.5-42.5 & 161-250 & $27^{+11}_{-11}$  & $19.7$  \\
10 GHz    & 32.5-42.5 & 181-250 & $27^{+12}_{-11}$ & 17.8  \\
\end{tabular}
\end{center}
\end{table}

\begin{table}[thp]
\caption{Likelihood results of the splits at 10 and 15 GHz}
\vspace{0.4cm}
\begin{center}
\begin{tabular}{lcccc}
\hline
Data & Dec. (\degg) & RA (\degg) & $\Delta T_\ell$ ($\mu$K) & $L_{max}$ \\
\hline
15 GHz A  & 27.5-45   & 161-230 & $31^{+9}_{-8}$ & $9.4\times 10^{4}$ \\
15 GHz B  & 27.5-45   & 161-230 & $27^{+9}_{-7}$ & $1.1\times 10^{ 4}$ \\
15 GHz (A-B)/2 & 27.5-45  & 161-230 & $\le 13$ & 1.2 \\
\hline
\hline
10 GHz A  & 32.5-42.5 & 161-230 & $<54$ & 1.1 \\
10 GHz B  & 32.5-42.5 & 161-230 & $<59$	& 1.6 \\
10 GHz (A-B)/2 & 32.5-42.5& 161-230 &$\le 34$ & 1.0 \\
\end{tabular}
\end{center}
\end{table}

\subsubsection{JOINT ANALYSIS}

The first model considered in the joint likelihood analysis of the 10
and 15 GHz data assumes that the same signal is present in the data at
both frequencies; this excludes a priori the possibility to have any
signal changing with frequency between 10 and 15 GHz.  For the reason
mentioned above, we limited the analysis to the section of data at
RA=$161\degg-230\degg$. Analyzing the five common strips at 10 and 15
GHz we obtain a signal $\Delta T_\ell=30^{+10}_{-8}$ $\mu$K with a
likelihood peak of $1.5\times 10^6$.  Comparing these with the results
of the analysis of each frequency separately, we see that the signal is
similar to that found at 10 and 15 GHz. This indicates the
compatibility between the data at both frequencies ({\it i.e.} the
signal present in both frequency has not only the same amplitude but
also comes from the same structures). However, this agreement is not
complete, as suggested by the shape of the likelihood function which is
broader than the likelihood function analyzing the 15 GHz alone.

Two observing frequencies potentially allows the separation of the CMB
and Galactic components. The likelihood analysis offers a natural
method to distinguish between signal with and without spectral
dependance (Gorski \et 1996). We assume that the signals in the data at
10 and 15 GHz are given by $\Delta T_\ell^{10,15}=\Delta
T_{\ell_{CMB}}+\Delta T_{\ell_{GAL}}^{10,15}$, where $\Delta
T_{\ell_{GAL}}^{15}=\Delta T_{\ell_{GAL}}^{10}\times (14.9/10.4)^\alpha
$, and $\alpha$ is the spectral index of the Galactic emission. The
model considers  that the CMB signal is characterized by a flat power
spectra along the window function of the experiment, while the power
spectra of the diffuse Galactic component is given by $\Delta
T_{\ell_{GAL}}\propto \ell^{-k}$ with $k=0.5$  (Gautier \et (1992),
Bersanelli \et 1996) which means a change in the Galactic contribution
by a factor $\sim \sqrt{3}$ along the window function of the
experiment. We have chosen values of $-2.15$ and $-3.0$ for the
spectral index of the Galactic component which correspond to two cases
in which this contribution is dominated by free-free and synchrotron
emission respectively. Analyzing the common strips at 10 and 15 GHz in
the RA range $161\degg-230\degg$ we obtain $\Delta
T_{\ell_{CMB}}=30^{+15}_{-11}$ $\mu$K and $\Delta T_{\ell_{GAL}}^{10} <28$
$\mu$K (68 \% C. L.) for the possible free-free component at 10 GHz;
similar results are obtained  assuming a synchrotron component. This value is our
best estimation of the CMB signal detected by the Tenerife experiments. There
is no detection of any Galactic component in any of the cases
considered. As a result the CMB signal has the same amplitude as in the
previous pure CMB models. As expected the uncertainty associated with
the CMB signal is larger now as a consequence of the new free parameter
in the analysis.

Fig. 7 presents this likelihood function. The results indicate the
consistency of the data at 10 and 15 GHz, and that the level of the Galactic
contamination is below the sensitivity of our data. 
Assuming a Harrison-Zeldovich spectrum for the primordial fluctuations,
the value corresponds to an expected quadrupole
$Q_{RMS-PS}=20^{+10}_{-7}$ $\mu$K (68 \% C.L.). This value includes the
uncertainties due to the instrumental error, the sample variance and
the possible existence of an undetected diffuse Galactic component.
This result is in agreement with $Q_{RMS-PS}=18\pm 1.6$ $\mu$K, the
value of the COBE DMR normalization (Bennett \et 1996), and implicitly
demonstrates the validity of the Harrison-Zeldovich model in the range
of multipoles covered by the COBE and Tenerife experiments ($l=2-30$).

\subsection{SPLITS OF THE DATA}

To check the consistency of the results we have split the data at 10
and 15 GHz into two halves $A$ and $B$ with similar sensitivities, and
compute the sum $(A+B)/2$ and the difference $(A-B)/2$. If both
halves are consistent they should contain the same signal while the
difference should contain just pure noise. We carried out the likelihood
analysis of these obtaining the results shown in Table~3. At 15 GHz this
analysis indicates the clear presence of signals with similar
amplitudes in both halves. The difference $(A-B)/2$ does not show the
presence of any signal, being compatible with noise. The likelihood
curves of the $A$, $B$, $(A+B)/2$ and $(A-B)/2$ data are plotted in
Figure~5. As expected at 10 GHz, the splits $A$ and $B$, and the difference
$(A-B)/2$ are compatible with noise, due to the lack of sensitivity and
the consistency between both halves respectively. 

An additional test has been done by computing the experimental
cross-correlation between the two halves $A$ and $B$ at 15 GHz, and
comparing it with the prediction for a given theoretical model. In our
case the beam response is not rotationally invariant due to the beam
switching tecnhique along lines of constant declination; this makes the
observed two-point correlation function depend on the separation in
RA-Dec. Computing it along the separation in RA we obtain the
correlation as a function of the separation in RA, $C(\theta_{ij})=\sum
_{ij} \Delta T_{i_1 j_1}\Delta T_{i_2 j_2} w_{i_1 j_1} w_{i_2 j_2}/\sum
_{ij} w_{i_1 j_1} w_{i_2 j_2}$. Figure~6 (top) shows this experimental
correlation function for the 15 GHz data. At the bottom of Fig.6 it is
represented the autocorrelation of the $(A-B)/2$ ({\it bottom}) showing
the compatibility between the data in the halves $A$ and $B$.  The
error bars in each case have been obtained by Monte Carlo techniques
(1000 simulations). The solid line is the expected correlation assuming
a signal given by a flat band power spectra with $\Delta T_\ell=30$
$\mu$K  in the angular region of sensitivity of our experiment. The
results obtained applying the same method to the 10 GHz are not
conclusive as a result of the larger noise in our data at this
frequency.

\section{CONCLUSIONS}

The Tenerife maps at 10 and 15 GHz  covering 5000 and 6500 square
degrees on the sky centred around  Dec.~+35\degg~have been presented;
from these we have selected $\sim 2000$ square degrees at high Galactic
latitude in which it is possible to analyze the properties of the CMB
signal.  The sensitivity achieved is enough to identify the most
intense features. The strongest radio sources in the data at 10 and 15
GHz have been identified at the level predicted. The data at 15 GHz
show the presence of a statistically very significant signal $\Delta
T_\ell=30^{+9}_{-7}$ $\mu$K; a similar analysis for the 10 GHz data
show the compatibility with this signal, but the poor sensitivity means
it is at the limit of detectability. A joint likelihood analysis of the
10 and 15 GHz data allowing for the presence of a CMB and a Galactic signal
gives values of $\Delta T_{\ell _{CMB}}=30^{+15}_{-11}$ $\mu$K and
$\Delta T_{\ell _{GAL}}^{10}=<28$ $\mu$K (68 \% C.L.). In the case of a
Harrison-Zeldovich spectra for the primordial fluctutions, the CMB
signal detected corresponds to $ Q_{RMS-PS}=20^{+10}_{-7}$ $\mu$K value
in good agreement with the normalization at large angular scales
provided by the COBE DMR data.  Both results support the standard
inflationary models.

The observations with the two radiometers continue in order to complete
the sky region between Dec.=+25\degg and Dec.=+45\degg~with uniform
sensitivity. The data for the radiometer at 33 GHz in the sky region
between Dec.=+35\degg~and Dec.=+42.5\degg~are now under analysis. This
extra spectral coverage will represent an important improvement to
separate the diffuse Galactic component from the CMB signal as was
shown in Hancock \et (1994). The CMB observations from the Teide
Observatory are being extended with several instruments working at
frequencies between 10 and 33 GHz and resolutions from a few degrees to
$10^{\prime}$. The first of the new instruments is a 33 GHz two-element
interferometer (Melhuish \et 1998) which has a resolution of 2\deg 5,
with full sine and cosine correlation in a 3 GHz bandwidth and which
has obtained its first measurements at Dec.=+41\degg$\,$ (Dicker \et
1998). Additionally a set of three instruments with several bands
centred at 10, 15 and 33 GHz which are being built at the IAC will
be installed during the year 1999. These instruments are designed to
obtain a map of flux and polarization on a region of the sky between
Dec.=+10\degg~and Dec.=+50\degg~with a resolution $\sim 1$\degg. The
collaboration between MRAO, NRAL and IAC continues with the Very Small
Array (VSA).  Operating at frequencies around 30 GHz at the Tenerife
site, the VSA will have the capability of imaging primordial CMB
structure to a sensitivity of 5 $\mu$K over the angular range
$10^{\prime}$ to 2\deg 5. This new set of instruments will provide a
large spectral and angular coverage which will make it possible to delineate
the CMB power spectra along the Sachs-Wolfe plateau and first few acoustic peaks, constraining some of the cosmological parameters at the 10 \% level.

\section*{Acknowledgments}{The Tenerife experiments are supported by
the UK Particle Physics and Astronomy Research Council, the European
Community Science programme contract SCI-ST920830, the Human Capital
and Mobility contract CHRXCT920079 and the Spanish DGES science
grant PB95-1132-C02.}

\clearpage

\subsection*{FIGURE CAPTIONS}

Figure 1. Mollweide projection of the sky showing the region observed by the Tenerife experiments.

Figure 2. The 10 and 15-GHz stacked scans along with their errorbars; ({\it left}) 15 GHz and ({\it right}) 10 GHz. The data have been binned in a 4\degg~bin in RA.

Figure 3. Maps at 10 and 15 GHz obtained by the Maximum Entropy Method.

Figure 4. Comparison bewteen the measurements at 10 and 15 GHz and the
estimation of the contribution of the point sources 3C 345, 3C39.25 and
3C286.

Figure 5. Likelihood curves obtained splitting the data at 15 GHz into
two halves $A$ and $B$, and computing the sum $(A+B)/2$ and the
difference $(A-B)/2$ respectively.

Figure 6. ({\it Top}) Croos-correlation of the splits $A$ and $B$ at 15
GHz and the theoretical auto-correlation function for a
Harrison-Zel'dovich spectrum, as convolved with the Tenerife beam and
normalised to $Q_{rms-ps} =20$ $\mu$K. ({\it Bottom}) Autocorrelation of
the $(A-B)/2$.

Figure 7. Likelihood  for the joint analysis of the 10 and 15 GHz as a
bivariate function of the amplitude of the Galactic and CMB signals.

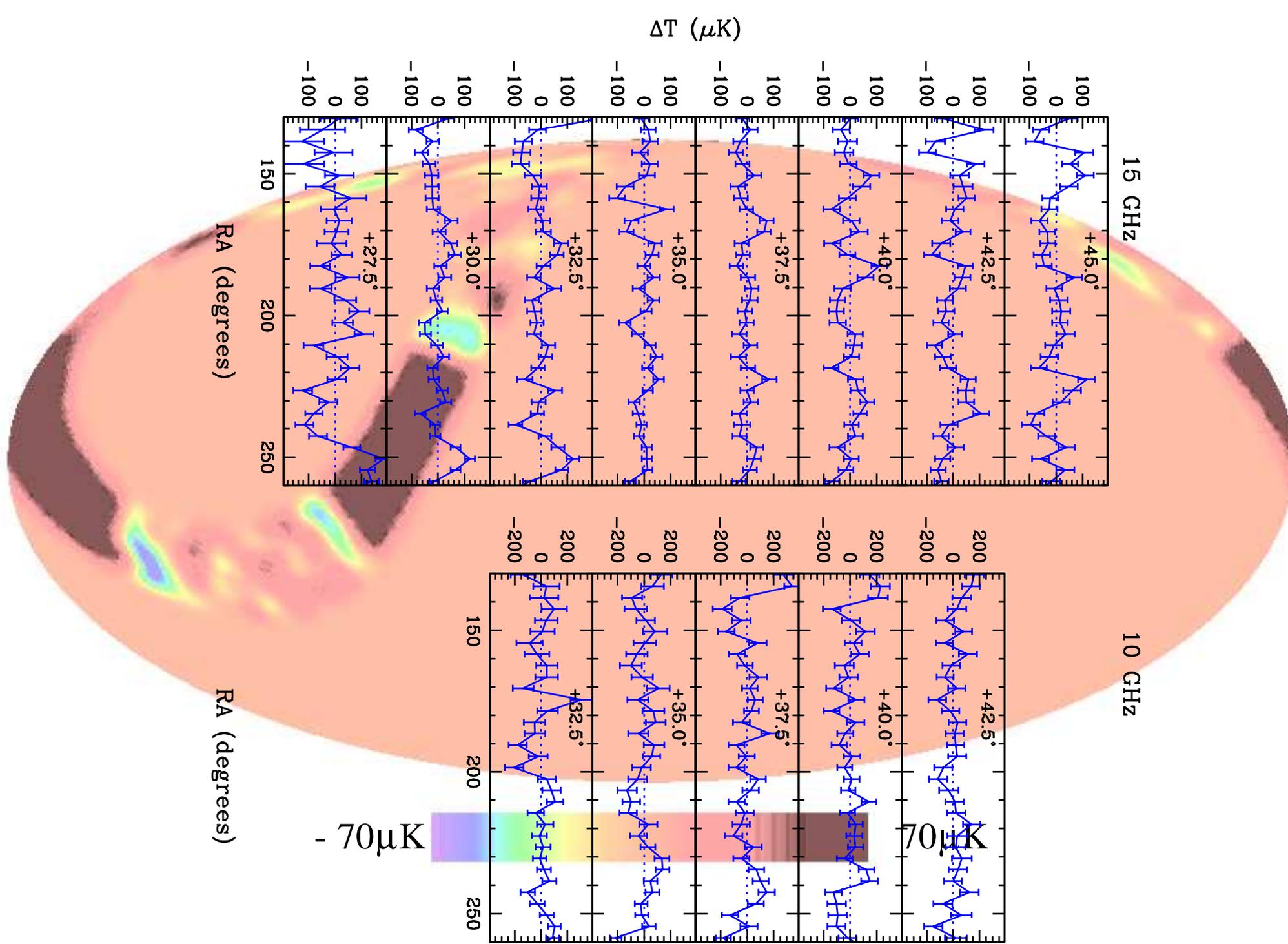
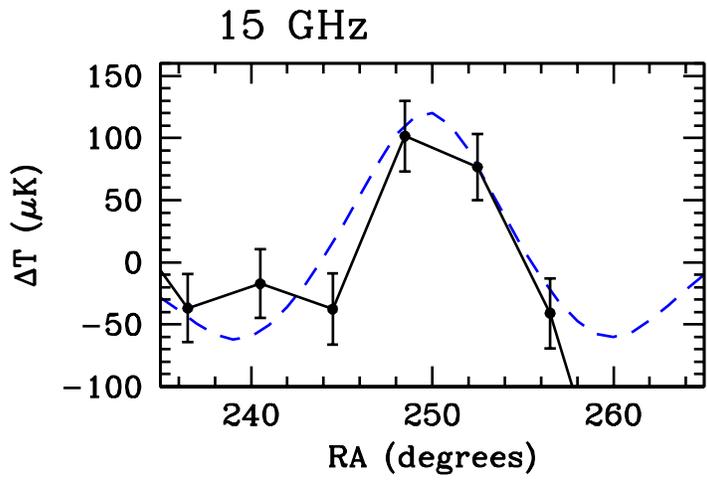
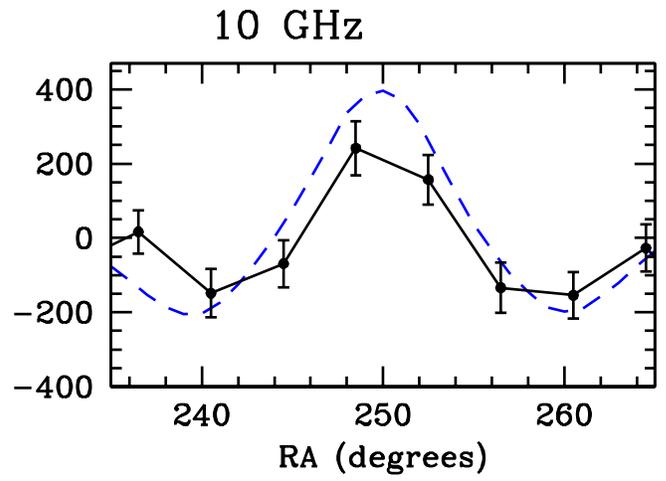
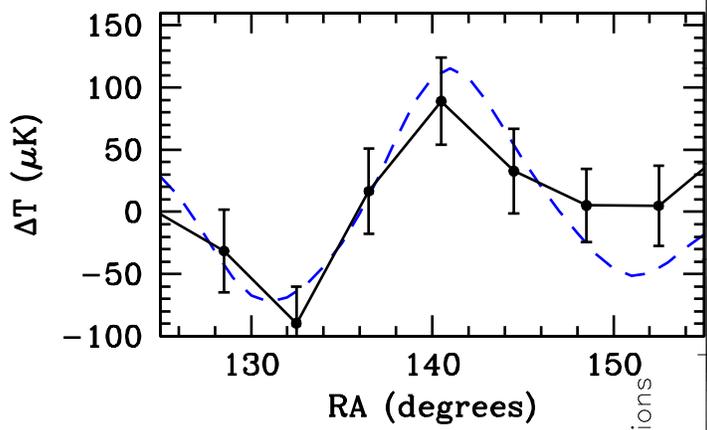
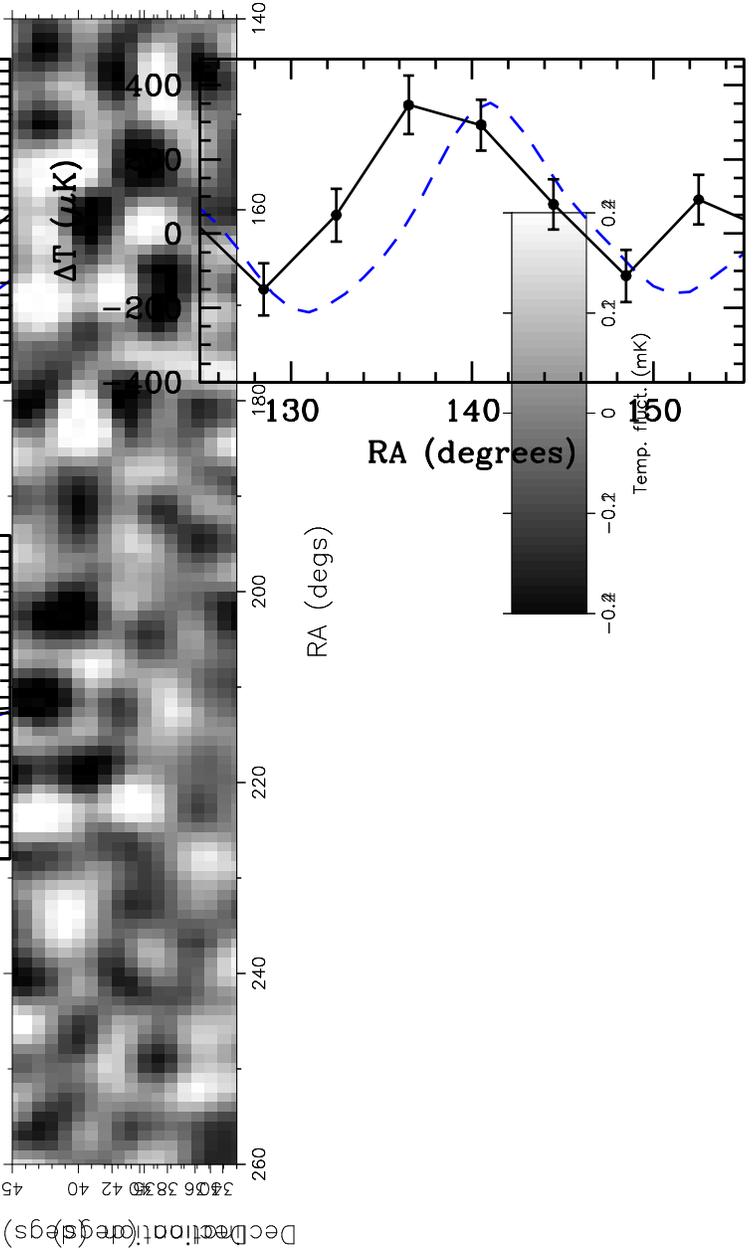
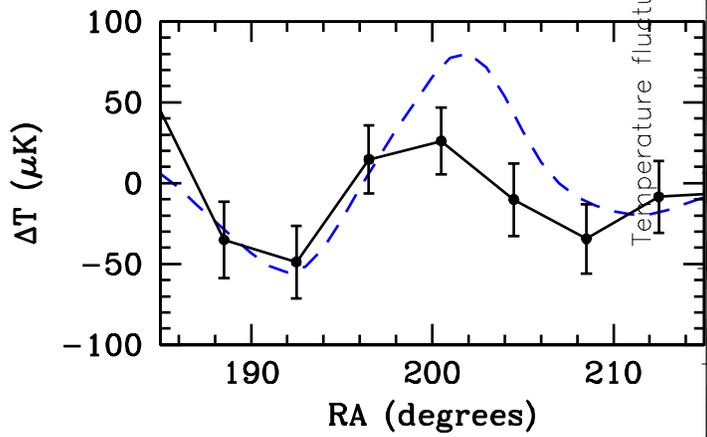
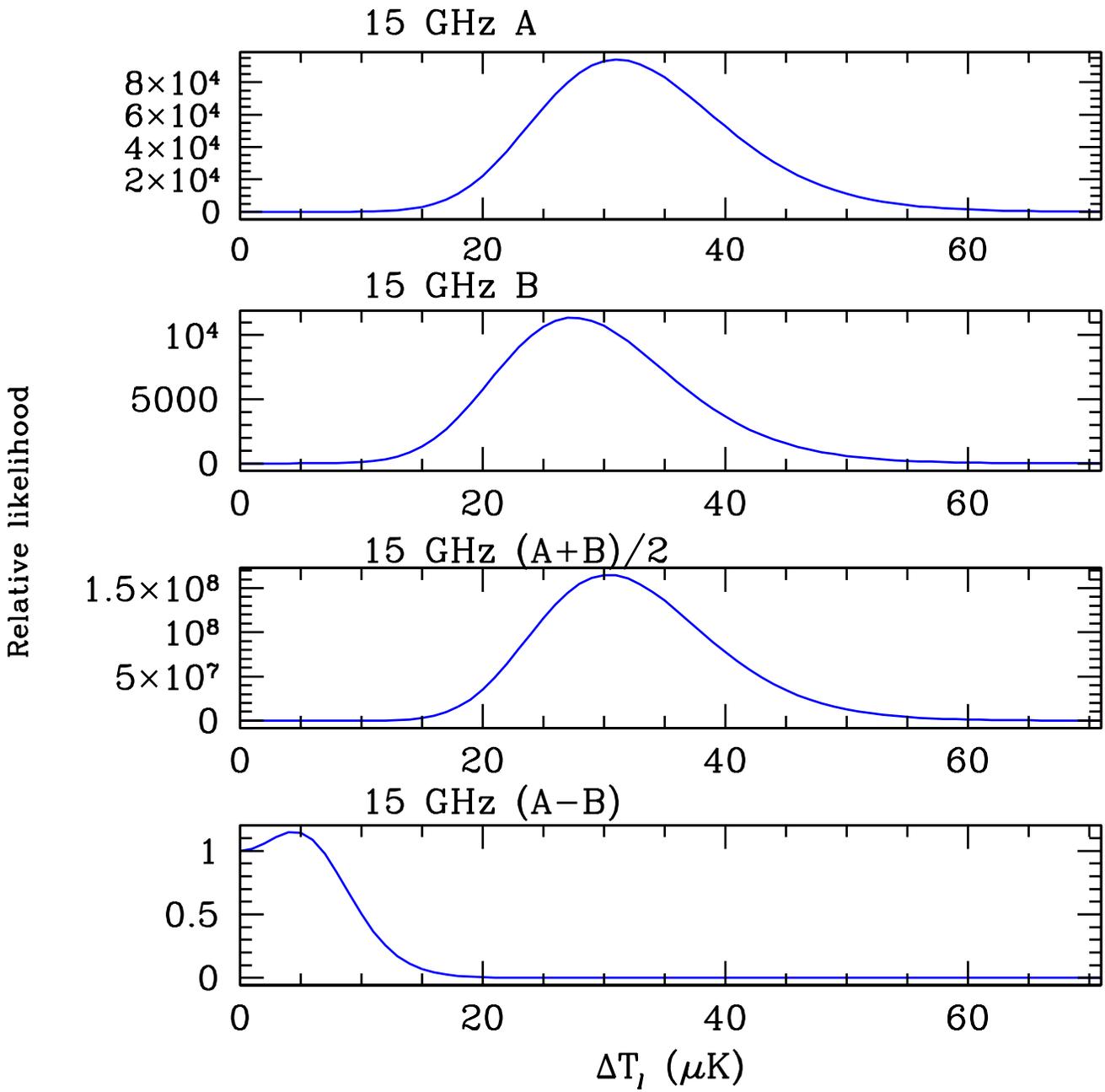
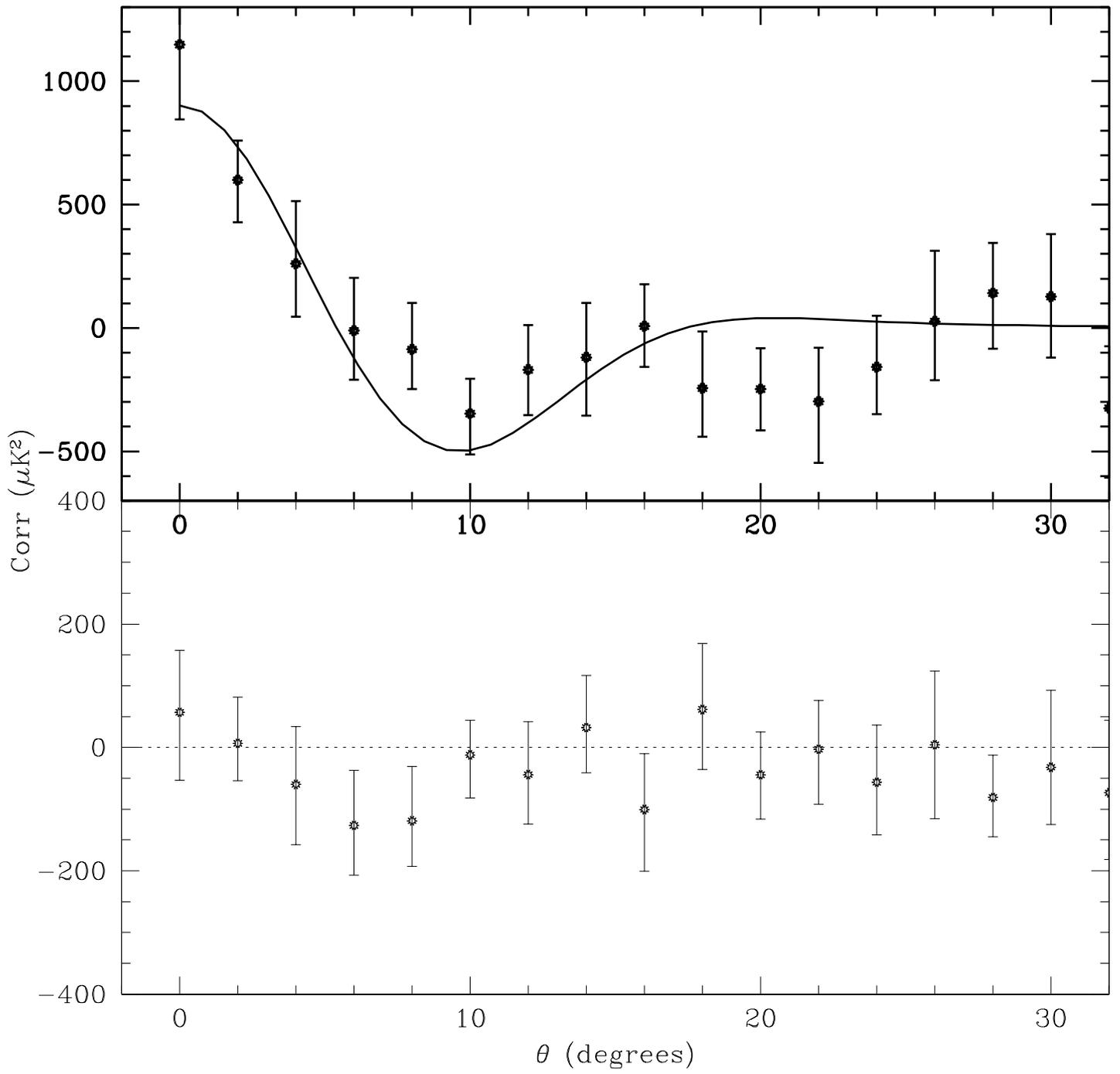
\end{document}